# Global gyrokinetic simulation of magnetic island induced ion temperature gradient turbulence in toroidal plasma

Jingchun Li[1,2*], J. Bao[3], Z. Lin[4], J. Q. Dong[5], Yong Liu[1], Y. R. Qu[2]

[1] College of Physics and Optoelectronic Engineering, Shenzhen University, Shenzhen, 518060, China
[2] Department of Earth and Space Sciences, Southern University of Science and Technology, Shenzhen, Guangdong 518055, China
[3] Beijing National Laboratory for Condensed Matter Physics and Laboratory of Soft Matter Physics, Institute of Physics, Chinese Academy of Sciences, Beijing 100190, China
[4] University of California, Irvine, CA 92697, United States of America
[5] Southwestern Institute of Physics, PO Box 432, Chengdu 610041, China
Email: jingchunli@pku.edu.cn

**Abstract**

The characteristics of ion temperature gradient (ITG) turbulence in the presence of a magnetic island are numerically investigated using a gyrokinetic model. We observe that in the absence of the usual ITG drive gradient, a solitary magnetic island alone can drive ITG instability. The magnetic island not only drives high-$n$ modes of ITG instability but also induces low-$n$ modes of vortex flow. Moreover, as the magnetic island width increases, the width of the vortex flow also increases. This implies that wider islands may more easily induce vortex flows. The study further indicates that the saturated amplitude and transport level of MI-induced ITG turbulence vary with different magnetic island widths. In general, larger magnetic islands enhance both particle and heat transport. When the magnetic island is of the order of 21 times the ion gyroradius ($21\rho_i$), the turbulence-driven transport level can reach the same level in cases where ITG is driven by pressure gradients.

## 1. INTRODUCTION

Magnetic islands are ubiquitous structures generated during the magnetic field reconnection process in magnetized fusion plasmas. In this universal phenomenon, unstable current sheets release magnetic energy through the reconnection process, leading the system to transition to a lower energy state by altering its topological structure—ultimately forming a magnetic island. This process is known as the tearing mode instability [1]. Due to the disruption of nested magnetic flux surfaces in the plasma, magnetic islands are considered a fundamental component of plasma transport. Currently, it is widely acknowledged that magnetic islands play a dual role in the confinement of fusion plasmas [2]. On one hand, the presence of magnetic islands induces a change in magnetic topology, creating fast radial transport channels through the X-point (reconnection point) in the random region surrounding the island separatrix. This results in a reduction in plasma confinement. The rapid growth of magnetic islands can also lead to plasma disruptions [3, 4]. On the other hand, magnetic islands are believed to contribute to the formation of internal transport barriers (ITBs) in the plasma. A stationary magnetic island generates strong shear flows near its separatrix, acting as a transport barrier and reducing the power threshold for ITB formation. This dual nature of magnetic islands makes their impact on turbulent transport a crucial and active area of research in plasma transport and confinement studies in fusion plasmas [5, 6, 7].

In recent years, with the development of advanced turbulence diagnostics and control methods for magnetic islands, detailed measurement results of magnetic islands and turbulence have been achieved on various devices such as DIII-D [8], KSTAR [9], HL-2A [10], LHD, J-TEXT [11], TJ-II [12], and W7-X. Simultaneously, a fast reciprocating Langmuir probe is utilized to study the spectral features of the dependence of geodesic acoustic modes (GAMs) and their nonlinear couplings with ambient turbulence on the magnetic island width. In EAST's H-mode discharges, the strongly coupled m/n = 1/1 internal modes triggering NTM magnetic islands have been observed, providing experimental evidence that instabilities generate magnetic islands through mode coupling [13,14]. Recent experimental studies also include investigations on the impact of magnetic islands on plasma flow and turbulence in the W7-X stellarator. The results indicate that the contribution of magnetic islands to flow is maximum at the island boundaries and approaches zero near the magnetic island

O-point. These observational findings share some similarities with results observed in other devices and nonlinear particle simulations[15].

Gyrokinetics has been extensively employed to investigate the interactions between magnetic islands and turbulence in toroidal fusion devices. Zarzoso et al. used the gyrokinetic code GKW to study the impact of rotating magnetic islands on ITG turbulence, revealing that in the nonlinear phase, magnetic islands can suppress turbulence levels within the island by up to 50% compared to the case without islands [16]. Similar results have been confirmed by local GENE simulations, showing that helical flows on both sides of the magnetic island exhibit increased shear strength [8]. In addition to early fluid studies on the impact of polarization currents on turbulent transport in small magnetic islands[17], a theoretical analysis of the influence of magnetic islands on ITG stability using gyrokinetic theory in slab geometry has been conducted. In this context, the flattening effect of magnetic islands is considered [18]. Recently, Zhang et al. investigated the influence of a magnetic island (MI) on electrostatic toroidal ITG mode. They found that when considering only the flattening effect of the magnetic island, the ITG mode can be stabilized compared to the case without the magnetic island. Simultaneously, the inclusion of MI-scale $\boldsymbol{E} \times \boldsymbol{B}$ flow enhances the effective drive of the ITG mode, providing theoretical evidence for the complex processes involving the influence of magnetic islands on ITG[19].

In addition to the research on the mutual influence between magnetic islands (MIs) and turbulence, another focus is on nonlinear mode coupling. Nonlinear mode coupling refers to the mutual evolution of low-n tearing modes of magnetic islands and smaller scale turbulence (higher mode numbers). This interaction has been extensively studied using fluid models [20], with a few exceptions [21]. It involves a two-way interaction, where not only is turbulence generated from magnetic islands, but there is also the phenomenon of turbulence nonlinearly coupling to generate magnetic islands. Regarding the latter, multiple numerical simulation studies have confirmed that small-scale interchange instabilities can transfer energy to the perturbation modes corresponding to magnetic islands through nonlinear three-wave interactions, thus exciting initial seed magnetic islands [22]. For the former, the special topology of magnetic islands can lead to a type of mode coupling associated with the distribution of rational surfaces. Fluid simulations by Wang et al. found that when the width of a magnetic island exceeds a certain level, a new radial non-local plasma Ion Temperature Gradient (ITG) mode is excited within its region, termed the Magnetic Island Induced ITG (MITG) mode [23]. However, Wang et al.'s work was conducted in a gyrokinetic fluid model, neglecting effects like Landau damping and the toroidal geometry of tokamaks. Importantly, whether magnetic islands alone can drive ITG turbulence remains unknown, as Wang et al.'s work still considers pressure gradients, which are the traditional driving source for ITG instability. Hence, providing an in-depth explanation of the nonlinear coupling between MIs and turbulence is crucial for advancing the understanding of multi-scale interactions in fusion plasmas, and this is the primary objective of the current study. In this letter, we present the global gyrokinetic analysis of the induction of ITG turbulence by a static magnetic island. The rest of the paper is organized as follows. The physical method is presented in section 2. The simulation results of the ITG turbulence generated by an MI and the transport characteristic under different island size are demonstrated in section 3. Finally, the conclusions are drawn in section 4.

## 2. SIMULATION APPROACH

We investigate the nonlinear impact of fixed magnetic islands on turbulent modes, adopting a method similar to that described in reference [24]. In our previous work, the focus was on the influence of magnetic islands on pressure-driven ion temperature gradient turbulence. In contrast, the current study addresses the impact of magnetic islands on the generation process of ITG turbulence. From a physical perspective, electrons have a significant thermal velocity and can rapidly respond to low-frequency perturbations, such as those arising from ITG instability, in the electrostatic field. Therefore, the primary response of electrons is adiabatic, with the adiabatic response predominantly involving the non-zonal flow component of the perturbed electric field on magnetic surfaces. However, distinguishing this zonal flow component concerning magnetic island topology is computational challenging. Consequently, many existing large-scale numerical simulation codes struggle to calculate the electron adiabatic response relative to magnetic island topology. To overcome this challenge, we employ drift kinetic electrons, allowing us to evolve the complete electron distribution function without the need to distinguish the adiabatic component. The introduction of drift kinetic electrons has the additional benefit of accurately reflecting the dynamical effects of electrons in the presence of magnetic islands, thus correctly describing the influence of trapped electrons on turbulence [25, 26]. This is particularly relevant for situations where the topology of magnetic islands is crucial.

Our model is grounded in first principles and employs gyrokinetic ions and drift kinetic electrons. In the specific simulations, we introduce a fixed magnetic island with toroidal and poloidal mode numbers m=2, n=1 into the background magnetic field. This magnetic island, located at r = 0.5a, corresponds to the rational surface position of $q$ = 2, and its width is adjusted manually by modifying a coefficient. The magnetic island is implemented through the perturbed magnetic vector potential $\delta A_\parallel$ = $-\nabla \times (R_0 B_0 \cos(2\theta - \phi))$, where $R_0$ and $a$ are the major and minor radii of the tokamak, respectively. The helicity of the magnetic island is m = 2, n = 1. Similar to the definition of magnetic island width in reference [16], the half-width $w$ of the magnetic island can be expressed in terms of the amplitude of the parallel component of the magnetic vector potential, safety factor q, magnetic shear $s=(r/q)(dq/dr)$, magnetic field strength B, and major radius $R_0$.

$$w^2 = \frac{2q\delta A_{\parallel 0} R_0}{Bs}$$

The actual width of the magnetic island in the simulations is determined by this definition.

## 3. SIMULATION RESULTS

The DKE model therefore represents an important advance in our understanding of plasma turbulence and its role in tokamak. The Cyclone base case-like equilibrium [24] is chosen in our simulations and the equilibrium safety factor ($q$) profile is shown in Fig. 1. Here we focus on the effect of $m/n=2/1$ magnetic island on the nonlinear evolution of bulk plasma turbulence. At the plasma axis, the ion and electron temperature are $T_e=T_i=2.22$ keV, the densities are $n_e=n_i=1.13\times10^{13}$cm$^{-3}$, the magnetic field strength is $B_0=2.01254\times10^4$ G, the dynamical plasma beta $\beta_e = 8\pi T_e/B_0^2=0.25\%$, and the ion cyclotron radius $\rho_i/R_0 = 2.86\times10^{-3}$. The $q=2$ rational surface (RS) is located at the center of the plasma $r=0.5a$, where the magnetic shear $s=(r/q)(dq/dr)=0.54$. Furthermore, the characteristic scale lengths of particle density and temperature are defined as $L_n=-(d\ln n/dr)^{-1}$, and $L_T=-(d\ln T/dr)^{-1}$, respectively. At the $q=2$ RS, we have set $R_0/L_{Ti}=R_0/L_{Te}=1.9$, $R_0/L_{ni}=R_0/L_{ne}=1.9$, and there exists no source for driving the turbulence.

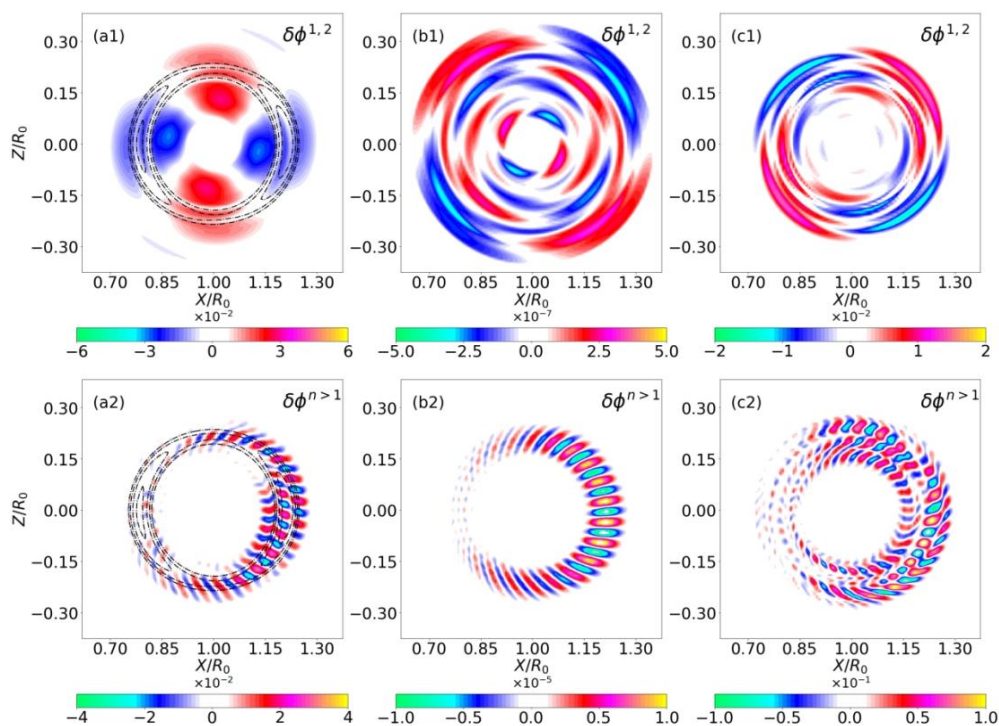

Fig. 1 Structure of the perturbed potential component $\delta\phi^{1,2}$ and $\delta\phi^{n>1}$ on the poloidal plane. (a1-a2):with MI and without ITG mode drive, (b1-b2): without ITG dirve and MI, (c1-c2): with ITG drive force and without MI.

Figure 1 shows the structure of the perturbed potential component $\delta\phi^{1,2}$ and

$\delta\phi^{n>1}$ on the poloidal plane, where the width of the magnetic island (MI) is approximately 14% of the minor radius (a). Clearly, when both the magnetic island and the pressure gradient are absent, the mode structure is essentially at the level of numerical noise (panel b). In the presence of only a pressure gradient, the pressure gradient can drive conventional ballooning mode structures indicative of the ion temperature gradient instability. Interestingly, in the absence of an ITG driving gradient, the magnetic island alone can drive ITG instability, and the amplitude at the same time can reach half of the amplitude of the gradient-driven ITG. For low-n mode structures, the MI-driven structures exhibit a better match with the electrostatic potential structure of tearing modes compared to the gradient-driven structures, which is what we refer to as vortex flow [24, 21].

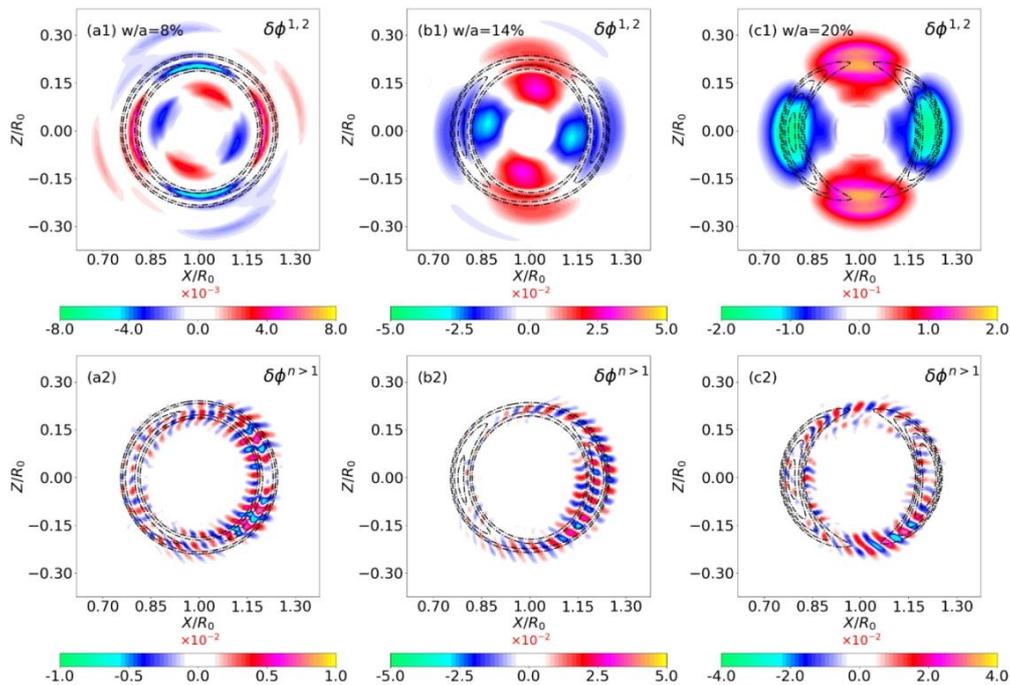

Fig. 2 Structure of the perturbed potential component $\delta\phi^{1,2}$ and $\delta\phi^{n>1}$ on the poloidal plane with different island size. (a1-a2) w/a= 8%, (b1-b2) w/a= 14%, (c1-c2) w/a= 20%.

Figure 2 depicts the high-*n* and low-*n* mode structures on the poloidal plane for different magnetic island widths. Whether it is vortex flow or the ballooning mode structures associated with ITG instability, larger MI widths correspond to larger amplitudes in both cases. This behavior mirrors the scenario in the presence of ITG gradients. As the MI width increases, the width of the vortex flow also increases, implying that wider islands may more easily drive vortex flows. This observation aligns with the analytical results previously presented by M. Leconte[27]. Starting from the extended Charney–Hasegawa–Mima equation with a non-axisymmetric equilibrium profile, M. Leconte's analysis reveals the coupled dynamics of drift-wave (DW) turbulence and flows. The saturation level of turbulence is determined by the damping rate of the island vortex-flow. Notably, this damping rate decreases with increasing island width (W) at a rate of approximately $1/w^2$. Consequently, it predicts a nonlinear threshold ($\gamma \sim 1/w^2$) for the formation of island vortex-flows, suggesting that wider magnetic islands may facilitate flow generation. This process aligns entirely with our findings in simulations.

Furthermore, it is observed in Fig.2 that the balloon-like structures weaken around the O-point, particularly in the case of small magnetic islands. Additionally, the ITG mode structures exhibit not only conventional balloon-like structures but also show the presence of structures at the inner magnetic island edge. This indicates that ITG structures are localized around the magnetic island, resulting in radially extended mode structures.

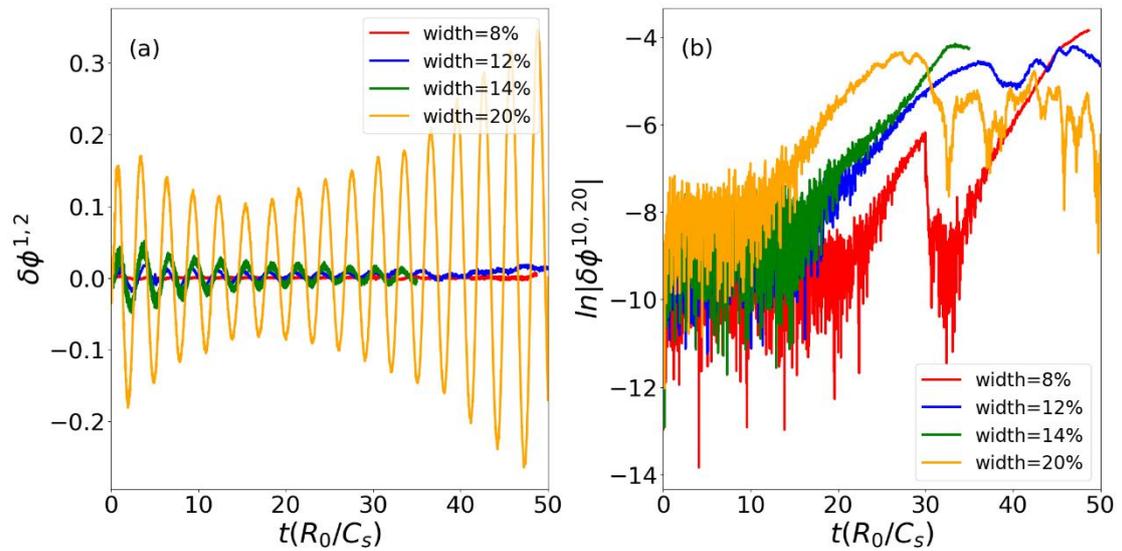

Fig. 3 Time evolution of each component of $\delta\phi^{1,2}$ and $\delta\phi^{n=10,20}$

Figure 3 illustrates the temporal evolution of low-n and high-n modes for different magnetic island (MI) widths. It is evident that larger magnetic islands result in higher final saturation amplitudes for low-$n$ modes. Additionally, it is noteworthy that despite varying sizes of magnetic islands, they exhibit the same oscillation frequency, approximately $\omega_{GAM} \approx 2.05 C_s/R_0$. This frequency closely resembles the frequency of the geodesic acoustic mode (GAM), $\omega_{GAM} \approx \sqrt{2\left(\frac{7}{4}\right)+1}\, C_s/R_0 = 2.12 C_s/R_0$, indicating synchronous behavior between this mode and the GAM. Analyzing the temporal evolution of high-$n$ modes, it is observed that different initial amplitudes exist for various magnetic island sizes (larger islands having larger amplitudes). However, the impact on their linear growth rate is relatively small, and the amplitudes tend to converge during saturation. This phenomenon is consistent with the results shown in panel *b* of Figure 2.

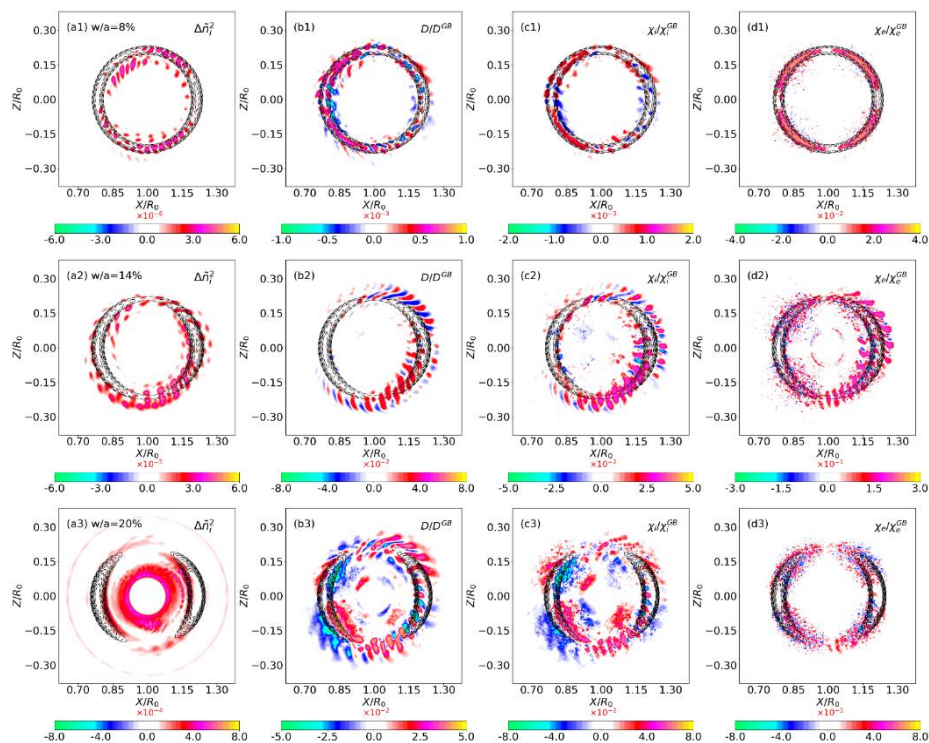

Fig. **4** The impact of MI width on MI driven turbulent intensity and transport in terms of relative amplitude, particle diffusion coefficient, ion and electron thermal conductive coefficients in gyro-Bohm units. (a1)-(d1): *w*/*a*=8%, (a2)-(d2): *w*/*a*=14%, (a3)-(d3): *w*/*a*=20%. The magnetic island structure is illustrated with the black dashed lines.

Our simulations indicate that the transport characteristics of turbulence solely driven by magnetic islands exhibit features distinct from turbulence transport driven by pressure gradients. Figure 4 illustrates the impact of magnetic islands on ITG turbulent transport under different magnetic island sizes. It is observed that, regardless of the magnetic island size, turbulent transport within the island is effectively suppressed. The saturation amplitude and transport level of ITG turbulence vary with different magnetic island widths. In general, larger magnetic islands enhance both particle and heat transport. Larger islands contribute to stronger transport near the X-point and in the external region, reducing plasma confinement levels, which is unfavorable for plasma stability. In comparison with Figure 7 in reference [24], we find that when w/$a$ =20%$a$ (approximately 21$\rho_i$ ), the turbulence transport induced by the magnetic island can reach a level similar to the pressure gradient-driven ITG transport under similar conditions. Therefore, ITG turbulence solely driven by magnetic islands can have a significant impact on plasma confinement and transport.

4. **Conclusions**

We conducted global nonlinear simulations using the electrostatic gyrokinetic equation, incorporating a static m/n=2/1 magnetic perturbation, to investigate the physical processes of magnetic island (MI) driving ion temperature gradient turbulence. For the first time, our simulations indicate that in the absence of the usual ITG driving gradients, a purely magnetic island can still drive ITG instability. The magnetic island not only drives ITG instability in high-n modes but also induces vortex flow in low-n modes, and as the magnetic island width increases, the vortex flow width also increases. This implies that wider islands may more easily drive vortex flows.

The characteristics of ITG turbulence transport induced by magnetic islands depend on the magnetic island width. Different widths of magnetic islands result in varied saturation amplitudes and transport levels of ITG turbulence. Specifically, larger magnetic islands enhance both particle and heat transport, thereby reducing plasma confinement performance. It is worth noting that, although kinetic calculations provide more reliable results compared to fluid models, the characteristics of turbulence presented here are still driven by static magnetic islands. Future studies requiring more accurate calculations should involve self-consistent dynamic magnetic islands.

**Acknowledgment**

This work was supported by the National Natural Science Foundation of China (11905109, 11905080 and 11947238), and Shenzhen Municipal Collaborative Innovation Technology Program - International Science and Technology (S&T) Cooperation Project (GJHZ20220913142609017). We would also like to thank Pro.f K. Ida, Z. W. Ma and G. S. Xu for helpful discussions.